\documentclass[times, twoside]{zHenriquesLab-StyleBioRxiv}
\usepackage{blindtext}
\usepackage{booktabs}
\usepackage{multirow}
\newcommand{\best}[1]{\textbf{#1}}

\leadauthor{Giedziun} 

\begin{document}

\title{Foundation Model-Driven Classification of Atypical Mitotic Figures with Domain-Aware Training Strategies}

\shorttitle{Foundation Models for MIDOG 2025}

\author[1,2]{Piotr Giedziun}
\author[2]{Jan Sołtysik}
\author[2]{Mateusz Górczany}
\author[1,2]{Norbert Ropiak}
\author[1,2]{Marcin Przymus}
\author[2]{Piotr Krajewski}
\author[2]{Jarosław Kwiecień}
\author[3,4]{Artur Bartczak}
\author[5]{Izabela Wasiak}
\author[6,7,8]{Mateusz Maniewski}
\affil[1]{Wrocław University of Science and Technology, Wroclaw, Poland}
\affil[2]{Cancer Center Sp. z o. o., Wroclaw, Poland}
\affil[3]{Hospital for Lung Diseases - Rebirth, Zakopane, Poland}
\affil[4]{Maria Sklodowska-Curie National Research Institute of Oncology, Krakow, Poland}
\affil[5]{Pathology Department, 10th Military Research Hospital in Bydgoszcz, Poland}
\affil[6]{Department of Tumor Pathology and Pathomorphology, Oncology Centre Prof. Franciszek Łukaszczyk Memorial Hospital, Bydgoszcz, Poland}
\affil[7]{Doctoral School of Medical and Health Sciences, Nicolaus Copernicus University in Toruń, Bydgoszcz, Poland}
\affil[8]{Department of Obstetrics, Gynaecology and Oncology, Collegium Medicum in Bydgoszcz, Nicolaus Copernicus University in Torun, Poland}

\maketitle

\begin{abstract}
We present a solution for the MIDOG 2025 Challenge Track~2, addressing binary classification of normal mitotic figures (NMFs) versus atypical mitotic figures (AMFs). The approach leverages pathology-specific foundation model H-optimus-0, selected based on recent cross-domain generalization benchmarks and our empirical testing, with Low-Rank Adaptation (LoRA) fine-tuning and MixUp augmentation. Implementation includes soft labels based on multi-expert consensus, hard negative mining, and adaptive focal loss, metric learning and domain adaptation. The method demonstrates both the promise and challenges of applying foundation models to this complex classification task, achieving reasonable performance in the preliminary evaluation phase.
\end{abstract}


\begin{corrauthor}
piotr.giedziun@cancercenter.ai
\end{corrauthor}

\section*{Introduction}

The MIDOG 2025 Challenge Track~2 focuses on differentiating normal mitotic figures (NMFs) from atypical mitotic figures (AMFs) in histopathological images, following the official structured challenge design \cite{midog25_challenge}. Atypical mitotic figures, characterized by abnormal chromosome distribution, are associated with more aggressive cancer behavior in breast cancer and increased likelihood of metastasis \cite{modpathol2022_amf}. The challenge presents four key obstacles: severe class imbalance, high morphological variability within classes, subtle differences between classes, and domain shifts across tumor types, species, and scanners. Task 2 uses balanced accuracy (BA) as an evaluation metric.

The combined dataset comprises 11{,}939 images from 503 whole slide images (WSIs) with severe class imbalance (14.8\% AMF vs 85.2\% NMF, ratio 1:5.74) distributed across 10 heterogeneous domains. Domain shift is substantial with AMF prevalence ranging from 7.4\% to 25.0\% across domains and variation in domain sizes, presenting significant challenges for generalization. 

\section*{Material and Methods}

\subsection*{Dataset and Data Splitting}

We combined three datasets to create a comprehensive training set:
\begin{itemize}
\item \textbf{MIDOG++ Dataset:} Comprehensive multi-domain dataset \cite{midogpp2023} providing the majority of training samples
\item \textbf{AMi-Br Dataset:} Human breast cancer mitoses with explicit atypical labels \cite{ami_br2025}; we removed the overlap with MIDOG++ (both derived from MIDOG21 dataset, files 001.tiff to 150.tiff) to prevent data duplication
\item \textbf{LUNG-MITO Dataset:} Lung-specific mitotic figure annotations expanding pulmonary coverage \cite{lung_mito2024}
\end{itemize}

After deduplication, the final combined dataset contains \textbf{11,939} images from \textbf{503} WSIs across \textbf{10} domains (combinations of 2 species, 7 tumor types, 4 scanners, and 4 institutions). The set is severely imbalanced with \textbf{1,771} AMFs (14.8\%) and \textbf{10,168} NMFs (85.2\%).

\paragraph{Data splitting strategy:}

We first performed Leave\mbox{-}One\mbox{-}Domain\mbox{-}Out (LODO) cross-validation to compare \emph{Virchow2} and \emph{H-optimus-0}, select LoRA hyperparameters, and determine the training horizon. For the final run, we pooled all data and created a single stratified split by the AMF label: \textbf{95\%} train and \textbf{5\%} monitor (used only for early stopping and checkpoint selection).

\subsection*{Augmentation Strategy Evaluation}

\begin{table*}[t]
\centering
\caption{Augmentation strategy comparison on MIDOG++ (5-fold CV). Baseline is absolute mean~$\pm$~std; other rows report $\Delta$ vs baseline (mean diff~$\pm$~std of that config).}
\begin{tabular}{lccccc}
\toprule
\textbf{Config} & \textbf{BA} & \textbf{F1} & \textbf{AUC} & \textbf{AMF Recall} & \textbf{NMF Recall} \\
\midrule
\textit{Baseline (absolute)} \\
baseline        & 0.602 $\pm$ 0.018 & 0.303 $\pm$ 0.040 & 0.652 $\pm$ 0.014 & 0.441 $\pm$ 0.073 & 0.763 $\pm$ 0.038 \\
\midrule
\multicolumn{6}{l}{\textit{Base augmentations ($\Delta$ vs baseline)}}\\
flip\&rotate    & +0.048 $\pm$ 0.011 & +0.057 $\pm$ 0.038 & +0.050 $\pm$ 0.014 & \best{+0.104 $\pm$ 0.032} & $-0.007$ $\pm$ 0.035 \\
color           & \best{+0.054 $\pm$ 0.027} & \best{+0.069 $\pm$ 0.052} & \best{+0.065 $\pm$ 0.033} & +0.080 $\pm$ 0.059 & +0.028 $\pm$ 0.025 \\
blur\&noise     & +0.001 $\pm$ 0.014 & +0.006 $\pm$ 0.037 & $-0.006$ $\pm$ 0.017 & $-0.021$ $\pm$ 0.013 & +0.024 $\pm$ 0.030 \\
distortion      & +0.030 $\pm$ 0.026 & +0.043 $\pm$ 0.048 & +0.026 $\pm$ 0.035 & +0.059 $\pm$ 0.094 & +0.000 $\pm$ 0.096 \\
cutout          & +0.019 $\pm$ 0.015 & +0.031 $\pm$ 0.037 & +0.030 $\pm$ 0.020 & $-0.002$ $\pm$ 0.023 & \best{+0.040 $\pm$ 0.052} \\
RASS       & +0.020 $\pm$ 0.020 & +0.031 $\pm$ 0.033 & +0.017 $\pm$ 0.014 & +0.002 $\pm$ 0.044 & +0.038 $\pm$ 0.031 \\
\midrule
\multicolumn{6}{l}{\textit{Composite pipelines ($\Delta$ vs baseline)}}\\
basic  & +0.023 $\pm$ 0.020 & +0.021 $\pm$ 0.039 & +0.029 $\pm$ 0.031 & +0.148 $\pm$ 0.057 & $-0.102$ $\pm$ 0.072 \\
medium & \best{+0.057 $\pm$ 0.024} & \best{+0.056 $\pm$ 0.044} & \best{+0.067 $\pm$ 0.037} & \best{+0.180 $\pm$ 0.054} & $-0.067$ $\pm$ 0.053 \\
heavy  & +0.024 $\pm$ 0.040 & +0.025 $\pm$ 0.060 & +0.005 $\pm$ 0.057 & +0.092 $\pm$ 0.056 & \best{$-0.044$ $\pm$ 0.025} \\
\midrule
\multicolumn{6}{l}{\textit{MixUp variants ($\Delta$ vs baseline)}}\\
standard mixup      & \best{+0.020 $\pm$ 0.017} & \best{+0.020 $\pm$ 0.029} & +0.007 $\pm$ 0.017 & +0.076 $\pm$ 0.011 & $-0.036$ $\pm$ 0.023 \\
strong mixup        & +0.015 $\pm$ 0.024 & +0.011 $\pm$ 0.047 & +0.003 $\pm$ 0.022 & \best{+0.124 $\pm$ 0.050} & $-0.095$ $\pm$ 0.006 \\
domain-aware mixup & +0.001 $\pm$ 0.008 & +0.006 $\pm$ 0.036 & \best{+0.009 $\pm$ 0.023} & $-0.024$ $\pm$ 0.039 & \best{+0.026 $\pm$ 0.026} \\
\bottomrule
\end{tabular}
\end{table*}

We conducted ablation studies to identify optimal augmentation strategies using EfficientNetV2-S on MIDOG++ with 5-fold stratified group-split cross-validation (5 epochs with early stopping). We evaluated a spectrum of spatial and photometric augmentations for AMF classification on our highly imbalanced, multi-domain dataset. 

Table 1 presents the comparative performance. Simple rotation and flipping already improved balanced accuracy from 0.602 to 0.650 by boosting AMF recall. The best performance was obtained with a medium composite pipeline that combines spatial transforms with restrained color jitter and light blur ($\Delta$BA \(+0.057 \pm 0.024\)), which outperformed baseline. Heavier augmentations (affine, cutout, strong noise/jitter) consistently favored the majority class (NMF) and reduced AMF sensitivity, leading to inferior balanced accuracy. Standard MixUp yielded a modest improvement ($\Delta$BA \(+0.020 \pm 0.017\)), whereas the domain-aware variant underperformed ($\Delta$BA \(+0.001 \pm 0.008\)). These results suggest that conservative, morphology-preserving augmentation is essential for this task, improving minority-class recall while maintaining adequate NMF performance.

We also evaluated Random Amplitude Spectrum Synthesis (RASS) \cite{qiao2024medicalimagesegmentationsinglesource}, a frequency-domain augmentation method designed to improve cross-domain robustness. However, in our experiments, RASS's balanced accuracy improvement ($\Delta$BA \(+0.020 \pm 0.017\)) came almost entirely from NMF recall ($\Delta$NMF +0.038) with negligible AMF recall change ($\Delta$AMF +0.002). While better than no augmentation, it underperformed simpler spatial and color transformations for detecting the minority class, leading us to exclude it from our final approach.

\subsection*{Model Architecture and Training}

\paragraph{Foundation Model:}
We evaluated two pathology foundation models - \emph{Virchow2} \cite{zimmermann2024virchow2scalingselfsupervisedmixed} and \emph{H\mbox{-}optimus\mbox{-}0} \cite{hoptimus0} - under Leave\mbox{-}One\mbox{-}Domain\mbox{-}Out (LODO) cross\mbox{-}validation. Based on superior cross\mbox{-}domain generalization in published benchmarks \cite{ammeling2025benchmarkingfoundationmodelsmitotic} and in our own experiments, we chose \emph{H\mbox{-}optimus\mbox{-}0} as the backbone. We fine\mbox{-}tuned it with parameter\mbox{-}efficient adapters (\textbf{LoRA}/PEFT) \cite{peft}, a strategy that has shown competitive gains for adapting pathology foundation models \cite{atnorm_br2025}. On top of the backbone, we added a compact MLP classifier head that produces a single logit. For domain awareness, we included a linear \textbf{domain head} with $|\mathcal{D}|$ outputs trained via a gradient\mbox{-}reversal layer.

\paragraph{Soft labels:} We calculated soft labels as the average of three expert annotations for each sample, transforming the binary classification into a soft-label problem that captures inter-annotator variability.

\paragraph{Hard Negative Mining:}
At each refresh, we run an evaluation pass over the training set and compute per-sample difficulty as the absolute error between predicted probability and label,
\(d_i=\lvert \hat{p}_i - y_i\rvert\) with \(\hat{p}_i=\sigma(f_\theta(x_i))\) (supports soft labels).
We mark the top \(30\%\) as \emph{hard} and double their sampling weight in the next epoch while keeping the rest in the pool (refreshed every epoch).

\paragraph{Adaptive focal loss:}
We use a sigmoid focal loss on logits with a positive-class prior to address the AMF:NMF imbalance. For a single logit–label pair $(z,y)$ with $y\in\{0,1\}$ (AMF$=1$), let $p=\sigma(z)$ and $p_t = y\,p + (1-y)(1-p)$. The loss is
\[
\ell(z,y) \;=\; (1 - p_t)^{\gamma}\;\ell_{\text{BCE-logits}}(z,y;\,w_{+}), 
\qquad \gamma=2.0,
\]
where $\ell_{\text{BCE-logits}}$ is the standard BCE-with-logits and $w_{+}$ (\texttt{pos\_weight}) is a dynamic positive-class weight computed from the observed class ratio (e.g., $w_{+}=N_{\text{NMF}}/N_{\text{AMF}}$, updated per epoch).

\paragraph{Metric learning components:}
We implemented a supervised contrastive regularizer to sharpen the embedding space. For each mini-batch, we extract $\ell_2$-normalized embeddings from the \emph{final hidden} layer and apply the \emph{Multi-Similarity} miner ($\varepsilon=0.1$) to select hard positives and hard negatives \cite{wang2020multisimilaritylossgeneralpair}. The mined pairs feed a \emph{Supervised Contrastive} loss with temperature $\tau=0.1$ \cite{khosla2021supervisedcontrastivelearning}. This metric term is weighted by $\lambda_{\text{con}}=0.5$ and added to the base focal loss. 

\paragraph{Sampling:}
We use a domain-aware weighted sampler. Each training example receives a weight that (i) increases for rarer classes, (ii) increases for underrepresented domains, and (iii) is doubled for items flagged as hard examples; all others retain their weight. Mini-batches are drawn with probability proportional to these weights, and the weights are refreshed after each hard-example mining pass.

\paragraph{Auxiliary loss:}
We experimented with auxiliary domain classification loss (weight=0.1) to encourage domain-invariant feature learning. 

\paragraph{Final Model:} Guided by the LODO findings that diversity aids generalization, our final model was trained on the pooled training split (95\%) with early stopping on the monitor split (5\%).

\section*{Results}

\subsection*{Cross-Domain Performance Evaluation}

Our model with LoRA fine-tuning was evaluated using 10-fold Leave-One-Domain-Out cross-validation, achieving a mean balanced accuracy of 0.851 $\pm$ 0.037 (AMF recall: 0.841 $\pm$ 0.088, NMF recall: 0.872 $\pm$ 0.038). The best performance was achieved on canine soft tissue sarcoma from Vienna (BA = 0.904, with AMF recall 0.929 and NMF recall 0.880), despite this being the smallest test domain with only 189 samples. In contrast, canine lymphoma proved most challenging (BA = 0.787) with particularly low AMF recall (0.634) despite being the largest test set with 3,959 samples.

Human domains showed consistent performance with balanced accuracies ranging from 0.837 to 0.874, while canine domains exhibited greater variability (0.787 to 0.904). Scanner variation within the same tumor type (human breast cancer on Hamamatsu XR vs S360) resulted in modest performance differences (BA: 0.860 vs 0.837). The model generally maintained better NMF recall than AMF recall across all domains, reflecting the inherent challenge of minority class detection despite class balancing strategies.

The results demonstrate strong generalization with mean balanced accuracy of 0.851 $\pm$ 0.037. 

\section*{Discussion}

Our foundation-model approach performed well overall, but \emph{out-of-domain (OOD)} generalization varied across laboratories, scanners, species, and tumor types; consistent with recent findings, \emph{LoRA}-adapting modern pathology foundation models nearly closes the OOD performance gap on unseen tumor domains \cite{ammeling2025benchmarkingfoundationmodelsmitotic}. In our augmentation experiments, we observed that moderate spatial and color transformations yielded the highest balanced accuracy (0.659 $\pm$ 0.024), while more aggressive augmentations corresponded with reduced minority class detection. MixUp variants, despite theoretical advantages, showed minimal improvement in our experiments, potentially related to the subtle morphological features that distinguish atypical figures.

The LODO evaluation revealed considerable performance variability across domains. The model on canine lymphoma yielded the lowest balanced accuracy (0.787) with notably reduced AMF recall (0.634), while canine soft tissue sarcoma from Vienna achieved the highest BA (0.904) despite having the smallest sample size (189 vs 3,959 samples). These observations suggest that factors beyond dataset size influenced model performance in our experiments, though the specific contributing factors remain unclear.

The metric learning framework we implemented addresses some of these domain-specific challenges by learning more discriminative feature representations that focus on hard examples within each domain, while the auxiliary domain loss does encourage learning of domain-invariant features.

Overall, pairing foundation models with domain-aware adaptation offers a promising route toward more consistent performance on unseen domains. The OOD gap observed in our LODO results are consistent with recent evidence for LoRA-adapted pathology FMs \cite{ammeling2025benchmarkingfoundationmodelsmitotic}.

\begin{acknowledgements}
We thank the MIDOG 2025 organizers for this valuable benchmark and the authors of MIDOG++, AMi-Br, and LUNG-MITO datasets for their contributions.
\end{acknowledgements}

\section*{Bibliography}
\bibliography{literature}

\end{document}